\documentclass[runningheads]{llncs}
\usepackage[T1]{fontenc}
\usepackage{graphicx}
\usepackage[backend=biber,style=ieee]{biblatex}
\begin{document}
\title{Preparing Students for AI-Driven Agile Development: A Project-Based AI Engineering Curriculum}
\titlerunning{Project-Based AI Engineering}
%

\author{Andreas Rausch\orcidID{0000-0002-6850-6409} \and
Stefan Wittek\orcidID{0009-0007-3877-625X} \and
Tobias Geger\orcidID{0009-0004-4469-534X} \and 
David Inkermann\orcidID{0000-0002-6587-816X}}
\authorrunning{A. Rausch et al.}
%
\institute{
Clausthal University of Technology, 38678 Clausthal-Zellerfeld, Germany\\
\email{\{andreas.rausch, stefan.wittek, thomas.tobias.marcello.geger, david.inkermann\}@tu-clausthal.de}}
\maketitle              
\begin{abstract}
Generative AI and agentic tools are reshaping agile software development, yet many engineering curricula still teach agile methods and AI competencies separately and largely lecture-based. This paper presents a project-based AI Engineering curriculum designed to prepare students for AI-driven agile development by integrating agile practices and AI-enabled engineering throughout the program.

We contribute (1) the curriculum concept and guiding principles, (2) a case study of interdisciplinary, AI-enabled agile student projects, and (3) early evidence from a mixed-methods evaluation. In our case study, second-semester bachelor students work in teams over seven two-week sprints on a realistic software product. AI tools are embedded into everyday agile engineering tasks—requirements clarification, backlog refinement, architectural reasoning, coding support, testing, and documentation—paired with reflection on human responsibility and quality.

Initial results indicate that the integrated approach supports hands-on competence development in AI-assisted engineering. Key observations highlight the need for agile teaching adaptations due to rapid tool evolution, the critical role of oral verification to ensure foundational learning. We close with lessons learned and recommendations for educators designing agile project-based curricula in the age of AI.

\end{abstract}
\section{Introduction}
Agile ways of working have become a de-facto standard across many engineering domains, increasingly beyond “classical” software projects. At the same time, generative AI and agentic tooling are reshaping how teams implement features, create tests, explore design alternatives, and plan work. As a result, core agile activities—such as backlog refinement, sprint planning, and retrospectives—are progressively influenced by AI assistance, while questions of quality, responsibility, and trust become more prominent when AI contributes to engineering artifacts.

Many engineering curricula, however, still treat agile methods and AI-enabled development as separate topics or as add-ons. Even when both are covered, they are often taught in lecture-centric formats that emphasize knowledge transfer over situated practice. This creates a gap: students may learn concepts and terminology, yet rarely experience how AI changes team dynamics, roles, collaboration patterns, and quality assurance in realistic agile projects.

We argue that closing this gap requires more than “AI in a lecture” or “Scrum in theory.” Effective preparation for AI-driven workplaces depends on authentic project settings with evolving requirements, real teamwork with explicit roles and responsibilities, continuous feedback cycles, and deliberate reflection on where human judgment and accountability remain essential. In other words, students need a learning environment in which agility and AI are practiced together as an integrated socio-technical capability.

To address this, we designed and iteratively refined a project-based engineering study program with an explicit AI engineering focus. The program is guided by eight principles: shaping digital transformation; technological innovation with quality and sustainability; human–AI collaboration; project-based learning; agility and reflection; practical relevance; entrepreneurship; and interdisciplinary teamwork. Within the curriculum, students repeatedly conduct AI-enabled agile projects across heterogeneous domains, applying agile practices end-to-end while using AI tools to support engineering tasks (e.g., ideation, coding support, test generation, documentation, stakeholder simulation). At the same time, they explicitly address what must remain a human responsibility, such as acceptance criteria, risk and ethics trade-offs, and accountability.


This paper contributes:
\begin{enumerate}
\item A study program concept for AI-enabled, project-based agile engineering education (structure, learning outcomes, guiding principles)
\item A teaching and project concept plus case study describing how AI tools are integrated into agile student projects (roles, events, artifacts, assessment and feedback)
\item An initial quantitative and qualitative evaluation with lessons learned and recommendations for preparing students for AI-driven workplaces
\end{enumerate}
  
The rest of the paper is structured as follows: Section 2 introduces the concept of our agile, project-based AI Engineering study program. Section 3 explains how we conduct interdisciplinary agile projects. Section 4 summarizes and discusses the initial evaluation results, combining qualitative and quantitative evidence. Finally, Section 5 concludes the paper and outlines directions for future work.  

\section{Designing AI-Integrated Agile Education: AI Engineering Study Program} \label{Study Program}

Teaching agile effectively requires learning formats that go beyond lecture-centric instruction: students need repeated, authentic project settings in which teamwork, evolving requirements, and continuous feedback are experienced rather than explained. At TU Clausthal, we have already operationalized this principle in the project-based study program Digital Technologies, where agile methods (e.g., Scrum), project management, creativity techniques, and teamwork are taught and internalized through practice in real projects \cite{digitec}, \cite{StenzelKuepperRauschetal.2025}.

With generative AI, the need for practice-oriented formats becomes even more pressing. LLMs and agentic AI increasingly support—and partly automate—engineering activities, but they also introduce new risks regarding quality, traceability, and responsible decision-making. Current curricula often still prepare students for a “workplace without AI,” creating a qualification gap. We therefore follow the same core educational hypothesis that proved successful for agile education: project-based learning is a powerful vehicle to develop AI competence in context, because it forces students to apply AI tools under real constraints, to reflect on limitations and failure modes, and to take responsibility for outcomes (“human oversight”). 

Building on these foundations, we developed the AI Engineering study program concept. Its key idea is to combine a shared core curriculum with discipline-specific specialization tracks. The core curriculum establishes a common baseline across all students—covering programming and software foundations, AI literacy and data-related skills, and agile development practices—so that students can collaborate in interdisciplinary settings. On top of this base, students choose an engineering focus (e.g., AI Software Engineering, AI Mechanical Engineering, AI Material Engineering), in which domain-specific engineering competencies are developed and repeatedly connected to the use of AI as an engineering tool (“AI as a tool for engineering,” not only “AI as a product”).
\begin{figure}[htbp]
    \centering
    \includegraphics[width=\linewidth]{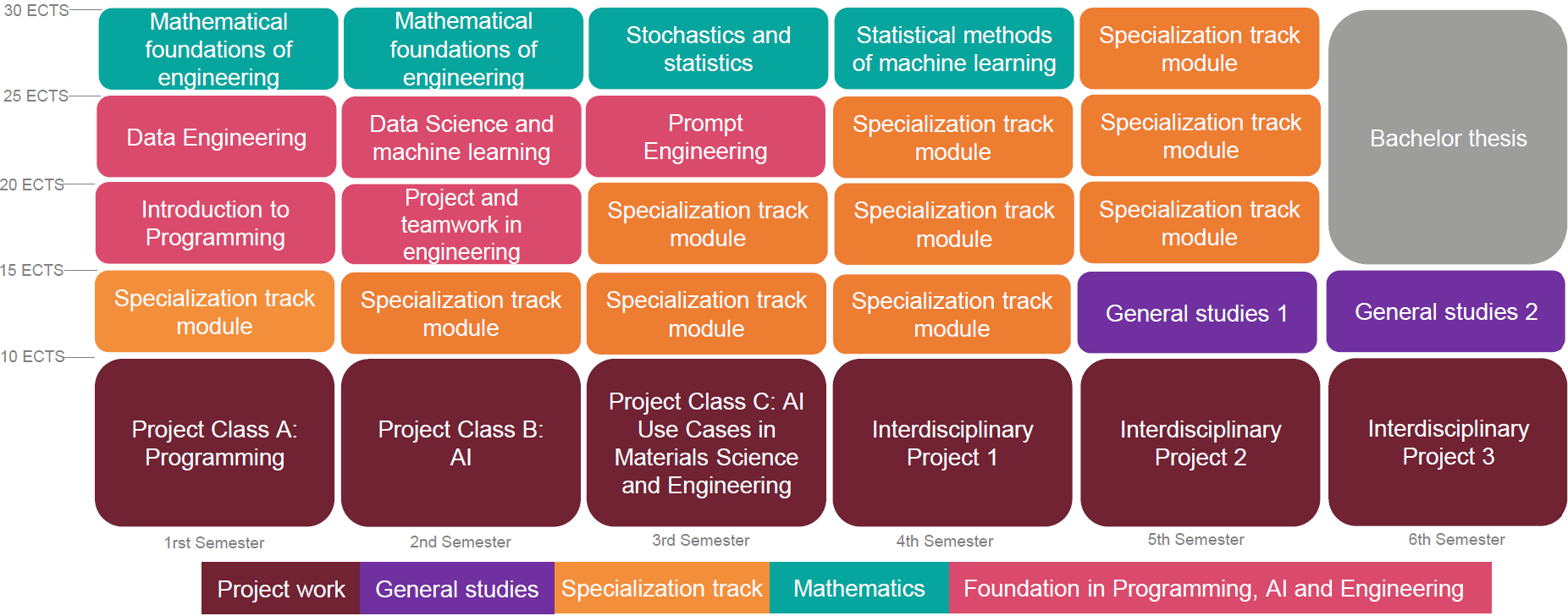}
    \caption{Model curricula of the bachelor program AI Engineering}
    \label{fig:curricula}
\end{figure}
A distinctive structural element of the program is that a substantial share of the workload is realized in project-oriented formats (approximately one third), where students apply agile methods and AI tools end-to-end on realistic problems—iteratively, in teams, and with continuous feedback. These projects are designed as authentic socio-technical learning settings: students experiment with AI-supported workflows, evaluate results critically, and learn how to embed AI usage into engineering quality assurance and accountability. The curriculum itself is treated as a learning, agile system that is iteratively improved with input from students, educators, industry, and societal stakeholders, and designed for transfer (e.g., reusable materials and modular extensions). 

The structure of the study program is summarized in the model curricula shown in Figure \ref{fig:curricula}: they make explicit (i) the shared core curriculum that establishes a common foundation in programming, mathematics, AI/data competencies, and agile ways of working, and (ii) the subsequent engineering specialization tracks (e.g., AI Software Engineering, AI Mechanical Engineering, AI Material Engineering) that deepen domain-specific knowledge while maintaining a consistent AI-augmented engineering perspective. Importantly, the figures also highlight the program’s strong project-based backbone: project classes and recurring interdisciplinary projects are distributed across the semesters to ensure that students repeatedly apply agile practices and AI tools in authentic team settings—first to build fundamental skills, then to develop and evaluate discipline-specific AI use cases, and finally to integrate them in larger interdisciplinary projects and the bachelor thesis.

\section{Case Study: AI-enabled Agile Projects} \label{Case Study}
The projects we present in this case study are conducted by second-semester bachelor's students. At this stage, the students possess basic programming skills in Python, knowledge of object-oriented programming in Java, and familiarity with fundamental development tools such as Git. Additionally, the foundations of UML-based modeling, Scrum-based development, and project management have been established in previous courses.

The projects in this case study are conducted by second-semester bachelor students. They enter with basic skills in Python, Java OOP, and Git, alongside foundations in UML modeling, Scrum, and project management established in prior courses.

The core objective is to transform these basics into comprehensive, AI-enabled team development capabilities. Students are expected to derive a domain model and consistent use-case diagrams, formulate epics and user stories, and master estimation techniques. They design a high-level architecture and adopt engineering practices including automated unit testing, clean code, and CI/CD pipelines. Throughout this process, they must demonstrate efficient and critical use of AI tools.
\begin{figure}
    \centering
    \includegraphics[width=1\linewidth]{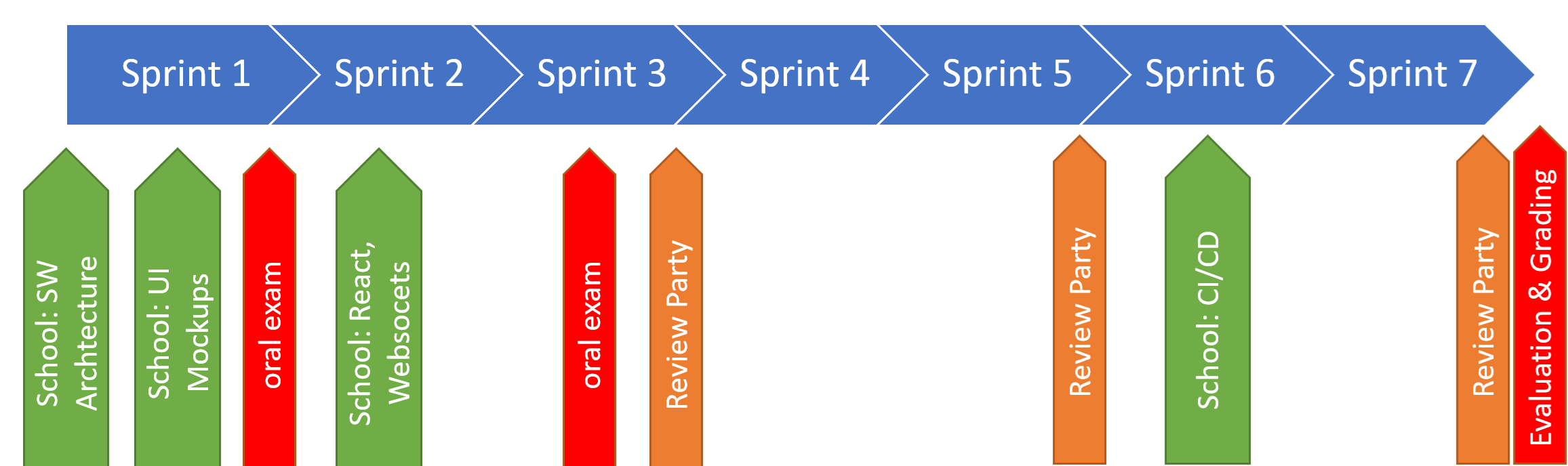}
    \caption{Time schedule of AI-enabled Agile Projects}
    \label{fig:schedule}
\end{figure}
Teams consist of 4--6 developers, a Product Owner (teaching assistant), and a Supervisor (professor). While students initially act as developers, they gradually assume Product Owner responsibilities, such as writing user stories. The technical goal is to develop a browser-based multiplayer board game \cite{hof}. The project follows the Scrum framework over seven two-week sprints. Standard events include bi-weekly Dailies, mid-sprint ``InBetween'' presentations, and Sprint Reviews. Additionally, cross-team ``Review Parties'' are held after sprints 3, 5, and 7 to present results to a wider audience.

Knowledge transfer occurs via ``Schools''—hybrid lecture-exercise units focused on technology (e.g., React, TypeScript) and architecture (e.g., client-server concepts). AI assistance is not a separate topic but is embedded into every School. Teachers demonstrate AI capabilities for feedback or code explanation, and homework assignments encourage students to use AI tools critically. Figure \ref{fig:schedule} shows the time schedule of a typical project along with its schools.

To ensure deep understanding despite AI code generation, students undergo individual 20-minute oral exams verifying their code and AI usage. Alongside the project, students maintain a presentation serving as living documentation of requirements, architecture, and process. The final grade is derived from the project repository and this presentation, adjusted by individual peer ratings from a guided final session.

\section{Discussion} \label{Discussion} 

This project structure was introduced in Summer Semester 2023 within the ``Digital Technologies'' program \cite{digitec} (established WS 2019). The WS 2022 cohort was the first to participate. Based on three completed runs, we identified three key lessons:

\begin{enumerate}
    \item \textbf{Agility:} The rapid emergence of production-ready AI tools required agile teaching, requiring to integrate topics like Copilot or agentic frameworks during ongoing projects.
    \item \textbf{Assesment:} Individual oral examinations are essential to ensure students grasp foundational concepts rather than blindly relying on AI.
    \item \textbf{Personification:} Students increasingly personify AI chatbots, occasionally describing ChatGPT as an additional team member.
\end{enumerate}

An initial quantitative evaluation indicates improved student performance starting with the WS 2022 cohort, with the average missing ECTS dropping from 25\% (WS 2021) to 3\% (WS 2024) (see Table~\ref{tab:ects_cohorts}).

\section{Conclusion} \label{Conclusion}
We presented a project-based AI Engineering curriculum that integrates agile practices with generative AI tools to prepare students for modern software development. Our case study demonstrates that authentic, interdisciplinary project settings significantly improve student performance and foster critical AI literacy. As AI tools evolve, future work will focus on adapting assessment models and exploring the long-term impact of AI teammates on agile team dynamics.

\begin{table}[h]
\centering
\caption{Average ECTS progression compared to curriculum, as of WS 2025}
\label{tab:ects_cohorts}
\begin{tabular}{|c|c|c|c|c|}
\hline
Intake & Avg. Current ECTS & Expected ECTS & Avg. Diff. & Avg. \% Diff. \\
\hline
WS 2021 & 135 & 180 & -45 & -25\% \\
\hline
WS 2022 & 158 & 180 & -22 & -12\% \\
\hline
WS 2023 & 115 & 120 & -5 & -4\% \\
\hline
WS 2024 & 58 & 60 & -2 & -3\% \\
\hline
\end{tabular}
\end{table}
%
%
%
\printbibliography
%




\end{document}